\documentclass[amsmath,superscriptaddress,amssymb,twocolumn,prl]{revtex4}
\usepackage{graphicx}
\usepackage{bm}
\usepackage{color}
\usepackage{dsfont}
\DeclareMathOperator{\Tr}{Tr}
\DeclareMathOperator{\Pf}{Pf}
\begin{document}

\title{Dislocation charges reveal two-dimensional topological crystalline invariants}
\author{Guido van Miert}
\affiliation{Institute for Theoretical Physics, Centre for Extreme Matter and Emergent Phenomena, Utrecht University, Princetonplein 5, 3584 CC Utrecht, The Netherlands}
\author{Carmine Ortix}
\affiliation{Institute for Theoretical Physics, Centre for Extreme Matter and Emergent Phenomena, Utrecht University, Princetonplein 5, 3584 CC Utrecht, The Netherlands}
\affiliation{Dipartimento di Fisica ``E. R. Caianiello", Universit\'a di Salerno, IT-84084 Fisciano, Italy}

\begin{abstract}

We identify a one-to-one correspondence between the charge localized around a dislocation characterized by a generic Burgers vector and the Berry phase associated with the electronic Bloch waves of two-dimensional crystalline insulators. Using this correspondence, we reveal a link between dislocation charges and the topological invariants of inversion and rotation symmetry-protected insulating phases both in the absence and in the presence of time-reversal symmetry.  Our findings demonstrate that dislocation charges can be used as generic probes of crystalline topologies. 
\end{abstract} 

\maketitle
\paragraph{Introduction --} The majority of 
topological phases of matter are characterized by anomalous surface states, whose presence or absence is linked to a topological index \cite{RMP82_3045,RMP83_1057}. 
The Chern number classifying quantum Hall insulators, for instance, counts the number of chiral states appearing at an isolated edge \cite{PRL49_405,PRL61_2015,PRL45_494}. 
Their anomaly clearly resides in the fact that it is impossible to have a one-dimensional (1D) crystal with a different number of right-moving and left-moving electronic channels. 
For topological phases protected by additional (non)spatial symmetries, surface states are anomalous 
since they do not fulfill the minimal requirements of the protecting symmetry. 
Time-reversal symmetry, for instance,  dictates that in a 1D system of spin one-half fermions, the Fermi energy $E_F$ must always intersect an even number of Kramers'  pairs, a condition that is clearly violated by the helical edge states of two-dimensional quantum spin-Hall insulators \cite{PRL95_226801,PRL95_146802,S314_1757,S318_766}. The Dirac cones appearing in three-dimensional topological (crystalline) insulators
are yet another example of these surface state anomalies \cite{PRL98_106803,PRB75_121306,PRB76_045302,NP5_438,PRB82_045122,NM12_422,PRL106_106802,NC3_982,ARCMP6_361, PRB94_045113}. The resolution of the paradox is simple: the Dirac cones, chiral or helical edge state appearing at the opposite edge always regularize the system and cancel the anomaly. However, revealing the anomaly of the electronic states at isolated surfaces with local probes gives immediate access to the topological invariant of the system \cite{N452_970,NP5_398,S325_178,S323_919,NP8_800}. 

As a matter of fact, point group symmetries can endow crystalline insulators with global topological invariants \cite{PRX7_041069}, which do not directly yield anomalous surface states \cite{PRX6_021015,PRB97_115143}. In the absence of time-reversal symmetry, for instance, two-dimensional inversion-symmetric crystals can be characterized by two $\mathbb{Z}_2$ topological invariants corresponding to the quotients of the parities of the Bloch waves at $M=(\pi,\pi)$ and $X_1=(\pi,0)$ and $M$ and $X_2=(0,\pi)$, respectively. With a proper choice of an indivisible crystal unit cell, these topological invariants 
give precise constraints on the quantized electronic contribution to the charge polarization~\cite{PRB86_115112,PRB88_085110}, and mandate the existence of 
protected modes in the entanglement spectrum~\cite{PRB83_245132}. 
The question that immediately arises is whether these signatures are the only diagnostic of the crystalline topology or if instead there exist different ``global" properties, {\it i.e.} contributed by all electrons in the system, revealing  the topological invariants of the system. In this Rapid Communication, we provide a positive answer to this question by showing that the total electronic charge trapped around a dislocation in a two-dimensional crystal always possesses a topological quantized contribution which directly pinpoints the bulk topological crystalline invariants of the system. Importantly, our electronic probe is able to diagnose also the crystalline topology of time-reversal symmetric systems, where the bulk electronic contribution to the charge polarization is trivialized, but where we show that crystalline topological invariants can be still defined.

\begin{figure}[tbp]
\centering
\includegraphics[width=\columnwidth]{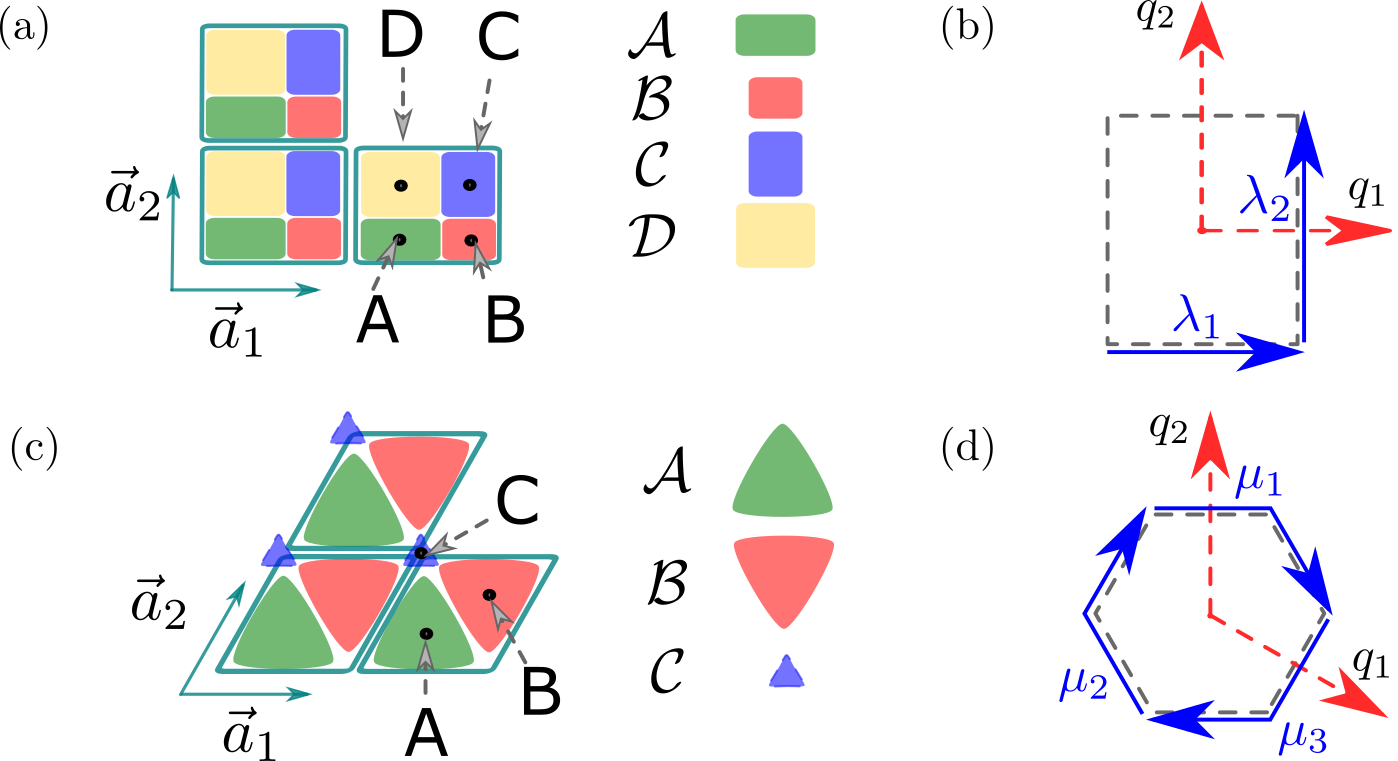}
\caption{Real-space and momentum space structure of inversion and rotationally invariants crystals. (a) Generic inversion-symmetric crystal. (b) Corresponding Brillouin zone with the two contours $\lambda_1$ and $\lambda_2$ along high-symmetry lines. (c) Generic threefold rotationally symmetric crystal. (d) Corresponding Brillouin zone, with the contours $\mu_1$, $\mu_2$ and $\mu_3$ along high-symmetry lines.}
\end{figure}

\paragraph{Crystalline topology of inversion- and two-fold rotation-symmetric crystalline insulators --}
We start out by defining the crystalline topology associated to inversion-symmetric two-dimensional insulators. The unit cell of a generic inversion-symmetric crystal can host four different inversion centers, thereby allowing for four inequivalent ``partitions" in the unit cell [c.f. Fig.~1(a)]. Let us now consider the inversion operator corresponding to the inversion center A of Fig.~1(a).  Under such inversion, an electron in the $\vec{i}$th unit-cell and belonging to partition $\mathcal{A}$ will be mapped to the $-\vec{i}$th unit-cell.  However, electrons belonging to partitions $\mathcal{B}$, $\mathcal{C}$, or $\mathcal{D}$ are sent to unit cells $-\vec{i}-(1,0)$, $-\vec{i}-(1,1)$, or $-\vec{i}-(0,1)$, respectively. As a consequence, we find that the Fourier transformed inversion operator acquires an explicit momentum dependence: $ \tilde{I}_A(\vec{q})=\textrm{diag}\left(
\tilde{I}_\mathcal{A},e^{-iq_1}\tilde{I}_\mathcal{B},e^{-i(q_1+q_2)}\tilde{I}_\mathcal{C},e^{-iq_2}\tilde{I}_\mathcal{D}\right)$. When performing the same analysis for the other inversion centers, one finds that all inversion operators are equal up to $\vec
{q}$-dependent phase factors, {\it i.e.} $\tilde{I}_A(\vec{q})=e^{-iq_1}\tilde{I}_B(\vec{q})=e^{-i(q_1+q_2)}\tilde{I}_C(\vec{q})=e^{-iq_2}\tilde{I}_D(\vec{q})$.
Henceforth, the topological invariants associated with inversion with respect to the different centers are all linked to each other. 
We emphasize that for crystals where the unit cell can be chosen to be centrosymmetric, the inversion operator looses any momentum dependence since one can identify a single partition in the unit cell. 

The inversion-symmetry protected
topological invariant can be identified by introducing the sewing matrix  $\mathcal{S}^{m,n}_{\tilde{I}_A}(\vec{q}):=\langle\Psi_m(-\vec{q})|\tilde{I}_A(\vec{q})|\Psi_n(\vec{q})\rangle$, with $|\Psi_m(\vec{q})\rangle$ denoting a Bloch wave with band index $m=1, \ldots, N_F$ and momentum $\vec{q}$, and where we chose the inversion center $A$ for convenience. This matrix tracks how the occupied states at $\vec{q}$ and $-\vec{q}=I\vec{q}$ are related by inversion symmetry. As long as the inversion operator commutes with the Hamiltonian,  $\mathcal{S}_{\tilde{I}_A}$ is a unitary matrix, and one can consider  the winding number of the corresponding determinant  along an arbitrary closed contour $C$: $
W_C(\mathcal{S}_{\tilde{I}_A}):=\frac{i}{2\pi}\int_C\mathrm{d}\vec{q}\cdot\nabla\log\det{\mathcal{S}_{\tilde{I}_A}(\vec{q})}\in\mathbb{Z}$. For a generic contour this integer is gauge dependent since under an arbitrary gauge transformation $|\Psi_m(\vec{q})\rangle\rightarrow\mathcal{U}_{m,n}(\vec{q})|\Psi_n(\vec{q})\rangle$, we have that $W_C(\mathcal{S}_{\tilde{I}_A})\rightarrow W_C(\mathcal{S}_{\tilde{I}_A})+W_C(\mathcal{U})+W_{I C}(\mathcal{U}^*)$. However, for contours that change orientation under inversion, i.e. $IC=-C$, the winding number can only change by a multiple of $2$, and as such it defines a $\mathbb{Z}_2$ invariant \cite{winding}.
In particular, we may consider the two contours $\lambda_1$ and $\lambda_2$, depicted in Fig.~1(b). The integrand for the winding number is even along these contours, 
and
therefore we can express both $\mathbb{Z}_2$-invariants in terms of the determinants at the inversion-invariant momenta 
\begin{align*}
\xi_{i=1,2,I_A}&:=e^{-i\pi W_{\lambda_i}(\mathcal{S}_{\tilde{I}_A})}=\frac{\det{\mathcal{S}_{\tilde{I}_A}(M)}}{\det{\mathcal{S}_{\tilde{I}_A}(X_i)}}.
\end{align*}
Inversion symmetry therefore endows two-dimensional crystalline insulators with a $\mathbb{Z}_2\times\mathbb{Z}_2$ topology \cite{PRB86_115112}.
Most importantly, these two $\mathbb{Z}_2$ numbers can be immediately linked to two Berry phases~\cite{PRB96_235130,PRB95_035421} $\gamma_{1/2}(q_{2/1})= \int_{0}^{2 \pi} d q_{1/2} \Tr\left[\mathcal{A}_{1/2}\left(\vec{q}\right)\right]$  obtained when viewing the two-dimensional system as a collection of momentum-dependent one-dimensional models, and where we introduced the Berry connection $\mathcal{A}^{m,n}_j(\vec{q})=\langle\Psi_m(\vec{q})|i\partial_{q_j}|\Psi_n(\vec{q})\rangle$ \cite{intercellular}. Using that the sum of the Berry connection at $\vec{q}$ and  $-\vec{q}$ can be expressed in terms of the sewing matrix $\mathcal{S}_{\tilde{I}_A}(\vec{q})$ and the bulk electronic charges $\rho_\mathcal{B}$, $\rho_\mathcal{C}$, and $\rho_\mathcal{D}$ in the partitions $\mathcal{B}$, $\mathcal{C}$, and $\mathcal{D}$, respectively, we find that  \cite{SM}
\begin{align}\label{eq:Berryinv}
 i\log{(\xi_{1/2,I_A})}=\frac{1}{2\pi}  \int_{0}^{2 \pi} \gamma_{1/2}(q_{2/1}) d q_{2/1} + \pi\left[\rho_\mathcal{B/D}+\rho_\mathcal{C}\right].
\end{align}
The equation above therefore provides a link between the topology of the ground state of the crystalline system, the geometric Berry phase and the bulk electronic charges of the insulator. 

We next show that such a link can be defined also when considering time-reversal symmetric systems of spin-one-half fermions. Kramer's theorem guarantees that inversion-symmetric crystals cannot be characterized topologically by the $\mathbb{Z}_2\times\mathbb{Z}_2$-invariants introduced above. This follows from the fact that at the time-reversal invariant momenta the determinant of the sewing matrices is always $1$. However, this does not 
preclude the existence of a different 
topology. In order to define new topological invariants, we consider the winding number of the determinants over half the contour $\lambda_{1(2)}$, i.e. from $X_{2(1)}$ to $M$. In general, this winding number is gauge dependent. However, if we impose a time-reversal symmetric gauge along these contours then both yield a $\mathbb{Z}_2$-invariant \cite{PRB96_235130,PRB94_165164}. We denote these winding numbers with $W^{1/2}_{\lambda_i}(\mathcal{S}_{\tilde{I}(\tilde{C}_{2,})_A})$ \cite{z2}. In inversion-symmetric insulators, we can simply express this invariant in terms of the eigenvalues of the inversion-operator for half of the Kramers partners at the high-symmetry points:
\begin{align*}
\chi_{i,\tilde{I}_A}&:=e^{-i\pi W^{1/2}_{\lambda_i}(\mathcal{S}_{\tilde{I}_A})}=\det{\mathcal{S}^I_{\tilde{I}_A}(M)}/\mathcal{S}^I_{\tilde{I}_A}(X_i).
\end{align*}
Here, $\mathcal{S}^I$ denotes the restriction of the sewing matrix to a single time-reversed ``channel". We point out that for two-fold rotation-symmetric insulators the invariants cannot be expressed in terms of symmetry eigenvalues due to the fact that $(\hat{C}_2\hat{T})^2=1$. This thus requires the full computation of the winding number over the contours indicated above. 
Similarly to the discussion above, 
we can express our crystalline topological invariants in terms of  geometric phases of the Bloch waves and the bulk electronic charges as 
\begin{align}\label{eq:pBerryinv}
i\log(\chi_{1/2,\tilde{X}_A})&= \pi q^I_{1/2} +\pi[\rho_{\mathcal{B}/\mathcal{D}}+\rho_\mathcal{C}]/2,
\end{align}
with $X=I$ or $C_2$, and where we introduced 
$$q^I_{1/2} = \frac{1}{2\pi^2}\int_0^\pi \gamma_{1/2}(q_{2/1}) \mathrm{d}q_{2/1}+\frac{1}{\pi}\gamma_{1/2}^I(\pi)-\frac{1}{2\pi}\gamma_{1/2}(\pi) $$
with $\gamma_{1/2}^I$ indicating the partial Berry phase, {\it i.e.} the Berry phase restricted to a single time-reversed channel \cite{SM}.
\paragraph{Three-fold rotational symmetric insulators--}
Before endowing the geometric phase terms appearing in Eq.~\ref{eq:Berryinv},~\ref{eq:pBerryinv} with a well-defined physical meaning, we generalize our results to crystals with a three-fold rotation symmetry. The unit cell of a $C_3$-symmetric crystal host $3$ rotation axes, which we label as A, B, and C, see Fig.~1(c). 
Precisely as for inversion-symmetric crystals, we can limit ourselves to consider the 
rotation axis A. One can show that the winding number of the sewing matrix $\mathcal{S}_{\tilde{C}_{3,A}}$ along the contour $\mu_1-\mu_3$ defines a $\mathbb{Z}_3$ invariant:
\begin{align*}
\xi_{C_{3,A}}:=e^{-i\frac{2\pi}{3}W_{\mu_1-\mu_3}(\mathcal{S}_{\tilde{C}_{3,A}})}=\frac{\det{\mathcal{S}_{\tilde{C}_{3,A}}(K^+)}}{\det{\mathcal{S}_{\tilde{C}_{3,A}}(K^-)}}
\end{align*}
with $K^\pm=\pm(2\pi/3,4\pi/3)$. Next, we use that the Berry connections at $\vec{q}$, $R\vec{q}$, and  $R^2\vec{q}$ are related to each other by the sewing matrix, with $R$ the rotation matrix corresponding to an anti-clockwise rotation of $2\pi/3$  \cite{rotation}. This allows us to write the equations linking the crystalline topological invariant to the geometric Berry phase and the bulk electronic charges as 
\begin{align}\label{eq:BerryC3}
i\log{(\xi_{C_{3,A}})} &=\frac{1}{2\pi}\int_{0}^{2 \pi} \gamma_{1}(q_{2}) d q_{2} + 2\pi[\rho_\mathcal{B}-\rho_\mathcal{C}]/3, \textrm{ and}\nonumber\\
i\log{(\xi_{C_{3,A}})} &=\frac{1}{2\pi}\int_{0}^{2 \pi} \gamma_{2}(q_{1}) d q_{1}  +2\pi[\rho_\mathcal{B}+2\rho_\mathcal{C}]/3.
\end{align}
Contrary to inversion- and two-fold rotational symmetric crystals,  this $\mathbb{Z}_3$-topology is not trivialized when accounting for time-reversal symmetric systems of spin-one-half fermions, for the very simple reason that $1+1\neq 0\textrm{ modulo }3$. In particular, since $K^+$ and $K^-$ are related by time-reversal symmetry, it follows that we can simplify the expression for $\xi_{C_{3,A}}$, by writing
$\xi_{C_{3,A}}=\det{\mathcal{S}_{\tilde{C}_{3,A}}\left(K^+\right)}^2=:\chi_{C_{3,A}}^2,$ 
with  $\chi_{{C}_{3,A}}$ representing a new topological invariant. Although this discrete quantity does not encode any new information, it can be directly related to the partial Berry phases and the bulk electronic charges in complete analogy with Eq.~\ref{eq:pBerryinv}. To show this, we can  impose a time-reversal symmetric gauge along the contours $\mu_i$ and compute the winding numbers $W^{1/2}_{\mu_{1}}$ and $W^{1/2}_{\mu_{3}}$,  from $X_2$ to $M$, and from $X_1$ to $M$, respectively. Then it follows \cite{SM}, that the winding number $W^{1/2}_{\mu_1-\mu_3}(\mathcal{S}_{\tilde{C}_{3,A}})$ is directly related to our novel topological invariant by $W_{\mu_1-\mu_3}^{1/2}(\mathcal{S}_{\tilde{C}_{3,A}})= 3 i\log(\chi_{C_{3,A}}) / (2 \pi)$. 
This, in turns,  allows us to express the bulk crystalline topological invariant as  
\begin{align}\label{eq:pBerryC3}
i\log{(\chi_{C_{3,A}})} &=\pi q_1^{I} +\pi[\rho_\mathcal{B}-\rho_\mathcal{C}]/3\textrm{ and }\nonumber\\
i\log{(\chi_{C_{3,A}})}&= \pi q_2^I +\pi[\rho_\mathcal{B}+2\rho_\mathcal{C}]/3.
\end{align}

\begin{figure}[tp]
\centering
\includegraphics[width=.475\textwidth]{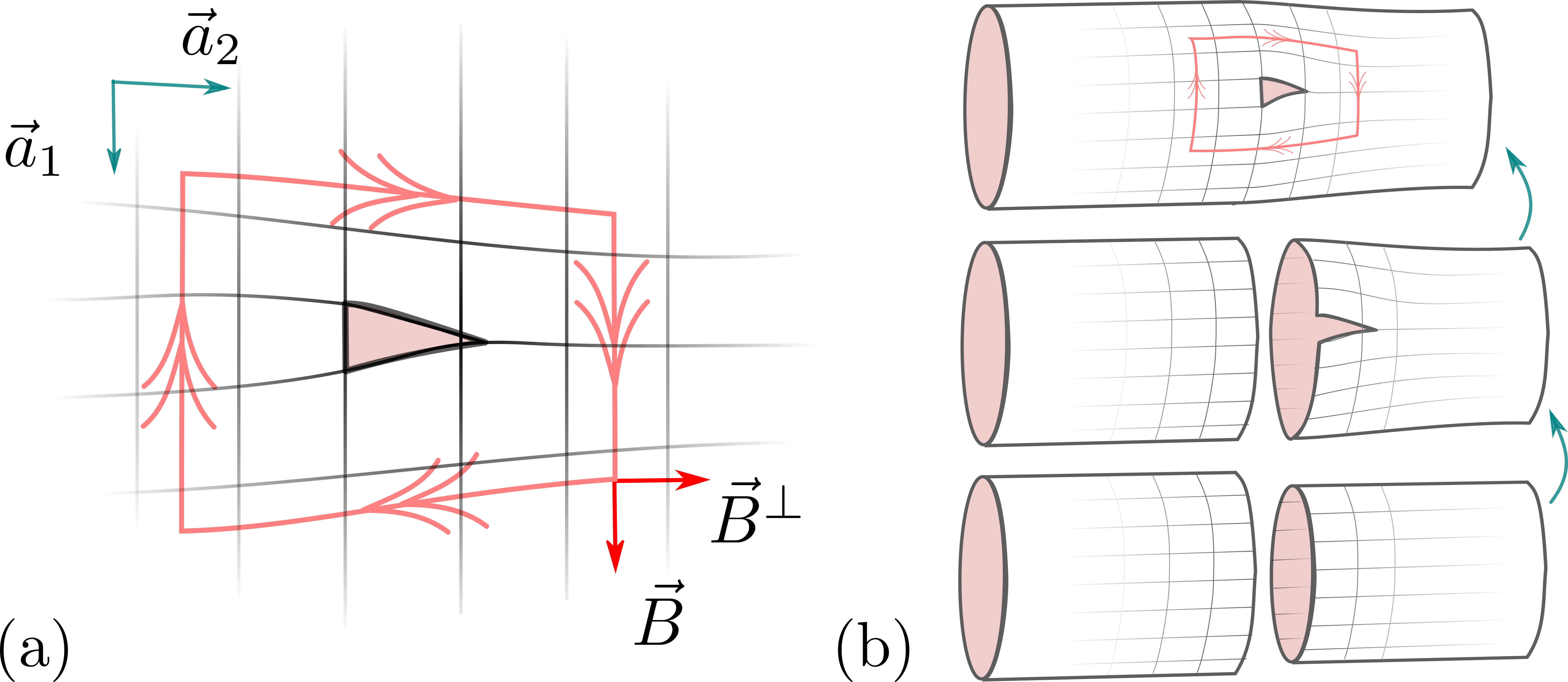}
\caption{(a) Schematic close-up of an edge dislocation. (b) Formation of a dislocation by gluing two cylinders. We find $\vec{B}=\vec{a}_1$, and $\vec{B}^\perp=\vec{b}_2/(2\pi\lVert\vec{a}_1\times\vec{a}_2\rVert)$. Here, $\vec{B}^\perp$ denotes the vector perpendicular to $\vec{B}$. The left (right) cylinder has $L_1$ ($L_1-1$) surface unit-cells.}
\end{figure}

\paragraph{Berry phase formulation of the dislocation charge --} We next show how the local electronic charges trapped around a dislocation can generally probe the crystalline topology we identified above. Dislocations are topological defects in  crystals. They can be associated with a topological invariant, the Burgers vector $\vec{B}$, which measures the difference between a defect-free and a distorted crystal.
 It can be defined as follows: One first traces out a loop surrounding the defect clockwise, and subsequently transfers this loop to a defect-free lattice. Due to the dislocation, the loop will fail to close. The vector that one needs to add to close the loop is the Burgers vector, see Fig.~2(a). The presence of this topological defect also produces a local distortion in the otherwise homogeneous charge distribution. For a conventional insulator, in fact, the charge per unit cell $\rho_{\vec{i}}$ in the vicinity of the dislocation core will deviate from its bulk value $N_F$, i.e. the number of occupied bands. 
Albeit the precise details of the charge distribution will depend on microscopic details, the dislocation charge $Q$, defined as the sum of the local charge deviations $\Delta\rho_{\vec{i}}=\rho_{\vec{i}}-N_F$, is a bulk quantity. Indeed, any local perturbation cannot alter the value of $Q$, since charge cannot flow away due to the insulating bulk. More specifically, and as shown below, it can be related to the Berry phase of the Bloch waves, and consequently used to probe the crystalline topological invariants. 
To derive such a relation, let us imagine that the dislocation is obtained by gluing two cylinders 
with different surface unit cells, 
as shown in Fig.~1. Since the total charge must be integer, we find that the dislocation charge (modulo 1) can be related to the edge charges of these cylinders via $Q=-Q_\textrm{L}-Q_\textrm{R}$, with $Q_\textrm{L(R)}$ the total left (right) edge charge. Both cylinders are periodic in the $\vec{a}_1$ direction, and thus can be viewed as a collection of 1D systems parametrized by $q_1=j2\pi/L_1$ and $q_1=j2\pi/(L_1-1)$, respectively. This, in turns, allows us to express $Q$ in terms of the corresponding Berry phases $\gamma_2(q_1)$, see Refs.~\cite{PRB96_235130,PRB95_035421}. 
In the $L_1\rightarrow\infty$ limit, we then find \cite{SM} that the dislocation charge is determined, modulo an integer, by
\begin{align}\label{eq:result1}
Q&=\frac{1}{4\pi^2}\int_0^{2\pi}\mathrm{d}q_1\gamma_2(q_1),
\end{align}
where we have used that $\gamma_2(q_1)$ depends continuously on $q_1$. Most importantly, when combined with Eq.~\ref{eq:Berryinv} the equation above implies that in the absence of time-reversal symmetry the dislocation charge of an inversion symmetric crystal contains a ``topologically" quantized contribution equal to $0$ or $1/2$ with the addition of a second unquantized contribution directly related to bulk electronic charge densities. The latter is absent for crystalline insulators which can be tiled with a bulk centrosymmetric unit cell, thereby implying a perfect quantization of the dislocation charge. The same holds true for three-fold rotational symmetric insulators , where the $\mathbb{Z}_3$ crystalline invariant implies that the topological contribution to the dislocation charge is quantized in multiples of $1/3$. We also point out that the expression for the charge trapped around a dislocation with Burgers vector $\vec{B}=\vec{a}_2$ allows for the general expression for the dislocation charge $Q= \vec{B}^\perp\cdot\vec{\gamma} / (2\pi)$ modulo $1$, where
$$\vec{\gamma}=\dfrac{\left[\int_0^{2 \pi}  \gamma_1(q_2) d q_2 \times  \vec{a}_1+\int_0^{2\pi} \gamma_2(q_1) d q_1 \times \vec{a}_2\right]}{( 2\pi \lVert \vec{a}_1\times\vec{a}_2\rVert)}$$ represents a Berry phase vector, and $\vec{B}^\perp$ the vector perpendicular to $\vec{B}$. 
The fact that the dislocation charge represents a genuine probe of the crystalline topology also applies when considering time-reversal symmetric crystals. In this case, the doubly degeneracy guaranteed by Kramers theorem implies that the dislocation charge is well-defined modulo $2$. We can account for this using the concept of partial Berry phase \cite{SM}, in terms of which the charge localized around a dislocation with Burgers vector $\vec{B}=\vec{a}_1$ simply reads $Q=q_2^I$, whereas for a generic dislocation $Q=\vec{B}^\perp\cdot\vec{q}^{\,I} $ modulo $2$, 
with $\vec{q}^{\,I}$ a partial Berry phase vector.  Therefore,  the dislocation charge of inversion-symmetric and time-reversal symmetric crystals contains a topological contribution quantized in odd integers, whereas for three-fold rotational symmetric crystals the discretized contribution comes in multiples of $2/3$. 

\begin{figure}[t]
\centering
\includegraphics[width=.475\textwidth]{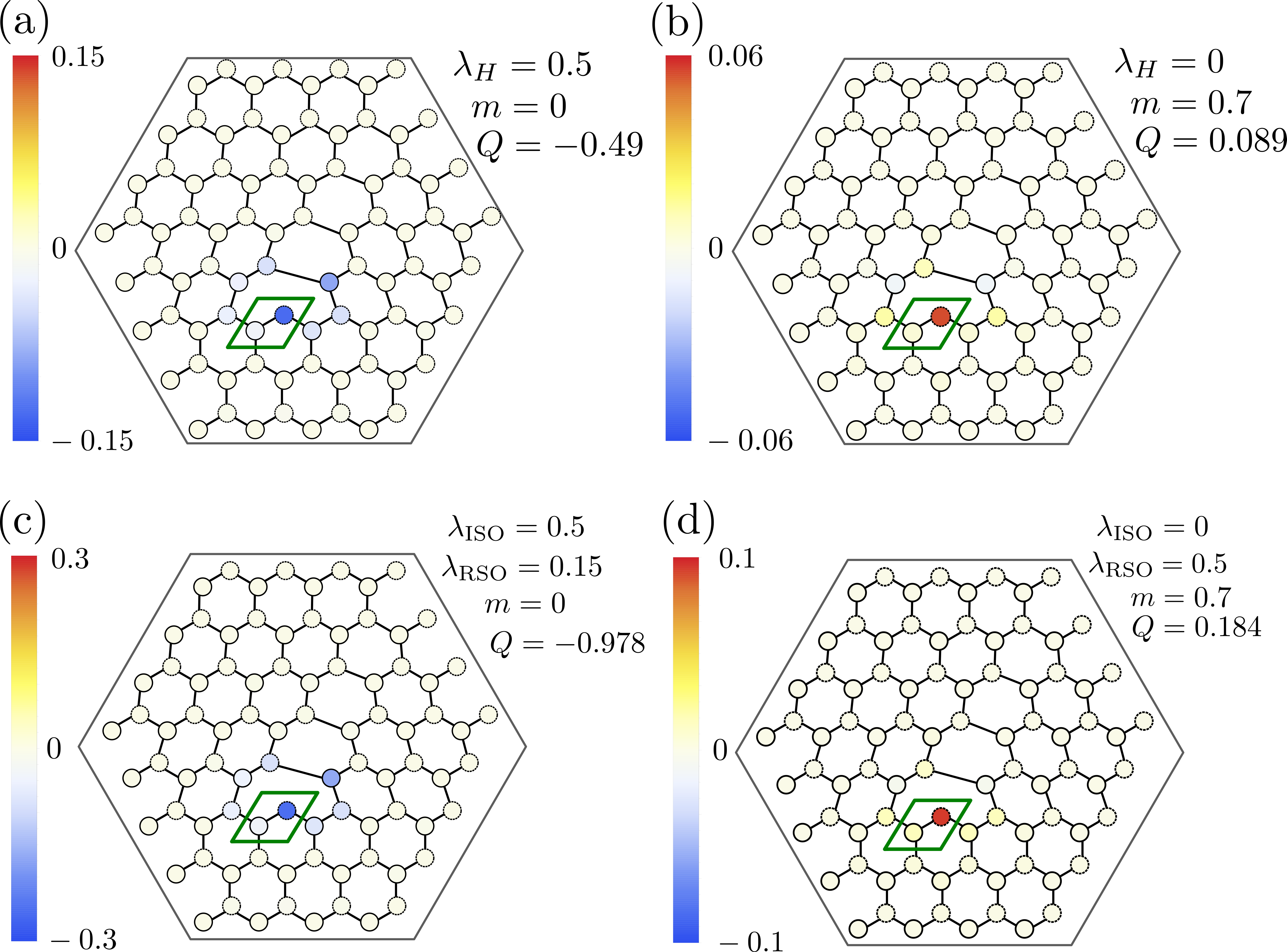}
\caption{Edge dislocation with Burgers vector $\vec{B}=\vec{a}_1$. The color quantifies the charge local charge deviation. The inset depicts the parameters that have been used, and $Q$ corresponds to the sum of the charge deviations within the hexagon. The upper-panels corresponds to a spinless honeycomb lattice, whereas the lower panels include spin-orbit coupling terms. In all cases we use $t=1$.}
\end{figure}

\paragraph{Honeycomb lattice model --} We check our findings against a honeycomb lattice model with a dislocation characterized by the Burgers vector $\vec{B}=\vec{a}_1$. We have first considered a model for spinless electrons that  hop between neighboring sites with a hopping parameter $t$, experience a staggered sub-lattice potential $\pm m$, and we also included a 
time-reversal symmetry breaking term corresponding to chiral orbital currents \cite{PRL61_2015}. In the tight-binding formulation the latter yields
imaginary next-nearest neighbor hopping processes. 
In momentum space, the bulk Hamiltonian globally reads:
\begin{align*}
\tilde{H}(\vec{q})&=\begin{pmatrix}
m+f(\vec{q})&-t(1+e^{-iq_1}+e^{-iq_2})\\
-t(1+e^{iq_1}+e^{iq_2})&-m-f(\vec{q})
\end{pmatrix}
\end{align*}
with $f(\vec{q})=2\lambda_H [\sin(q_1)- \sin(q_2) + \sin(q_2 - q_1)]$. 
We first set 
$m=0$, in which case the system is inversion-symmetric, 
with the inversion operator given by  the first Pauli matrix $\tau_1$. Its matrix representation is momentum independent since the unit cell is globally inversion-symmetric.  At half-filling we find $\xi_{1,I_A}=\xi_{2,I_A}=-1$ and hence for the dislocation depicted in Fig.~3(a), Eq.~\eqref{eq:result1} and Eq.~\eqref{eq:Berryinv} predict a dislocation charge $Q=1/2$. We have numerically verified this result by computing the local charge around the dislocation as indicated in Fig.~3(a).
 We have also analyzed the effect of the inversion-symmetry breaking term reducing the crystal symmetry to $C_3$. 
 The rotation operator has the matrix representation $\tilde{C}_{3,A}(\vec{q})=\textrm{diag}(1,e^{-iq_2})$, and acquires a momentum dependence since  the unit-cell is not rotation-symmetric. At half-filling the $\mathbb{Z}_3$ invariant is given by
\begin{align*}
\xi_{C_{3,A}}&=\begin{cases}1 &\textrm{ if }m<-3\sqrt{3}|\lambda_H|\\
e^{i2\pi/3} &\textrm{ if }-3\sqrt{3}|\lambda_H|<m<3\sqrt{3}|\lambda_H|\\
e^{-i2\pi/3} &\textrm{ if }m>3\sqrt{3}|\lambda_H|
\end{cases}
\end{align*}
Let us analyze the case $\lambda_H=0$ and $m=0.7t$. We have numerically calculated the charge in the $\mathcal{B}$ partition, which is given by $\rho_\mathcal{B}=0.73$. Hence, using Eqs.~\eqref{eq:result1} and \eqref{eq:BerryC3} we predict $Q=1/3-0.73/3=0.09$, which is in perfect agreement with the numerical result shown in Fig.~3(b).

Finally, we have considered a honeycomb lattice model for spin-one-half fermions where time-reversal symmetry is preserved. In particular, we included both intrinsic and Rashba spin-orbit coupling \cite{PRL95_226801}. The former corresponds to spin-dependent chiral orbital currents, whereas the latter leads to nearest-neighbor hoppings that flip the spin \cite{SM}. It is easily verified that both terms respect the  $C_3$ and $C_2$ -symmetry. For the latter, the corresponding symmetry operators reads:
\begin{align*}
\tilde{C}_{2,A}(\vec{q})&=\tau_1\otimes i \sigma_3 \textrm{ and } \tilde{C}_{3,A}(\vec{q})=\textrm{diag}\left(1,e^{-iq_2}\right)\otimes e^{i\frac{\pi\sigma_3}{3}}
\end{align*}
For the example depicted in Fig.~3(c) we find $\chi_{1,C_{2,A}}=\chi_{2,C_{2,A}}=-1$, which yields $Q=1 \textrm{ modulo } 2$. This result is numerically confirmed. In Fig.~3(d) we have depicted a spinful $C_3$-symmetric crystal. Using these parameters we find $\chi_{C_{3,A}}=e^{-i2\pi/3}$. Moreover, we find that $\rho_\mathcal{B}=1.44$. Therefore, we predict $Q=2/3-\rho_\mathcal{B}/3=0.187$. Again, this is in perfect agreement with the numerical result shown in Fig.~3(d).

\paragraph{Conclusions --}
To conclude, we have shown that the fractional part of the dislocation charge is a bulk property that can be linked to the Berry phase. In time-reversal symmetric systems, one can link the dislocation charge to the partial Berry phase, and microscopic details can only change the charge by multiples of $2$. We have also shown that in inversion- and rotation symmetric insulators, one can express the (partial) Berry phases in terms of topological crystalline invariants. 
Topological defects generally allow to identify non-trivial states of matter both in superconductors~\cite{PRL111_047006,PRB89_224503} and in insulators~\cite{NP5_298,S357_61,PRB96_245115,PRB89_161117,PRL108_106403,NP9_98}. Our results establish on firm grounds that dislocation charges uniquely probe topological crystalline insulators protected only by spatial symmetries. 

C.O. acknowledges support from a VIDI grant (Project 680-47-543) financed by the Netherlands Organization for Scientific Research (NWO). This work is part of the research programme of the Foundation for Fundamental Research on Matter (FOM), which is part of the Netherlands Organization for Scientific Research (NWO).

\appendix
\begin{widetext}
\section{Supplemental Material}

\section{Notation}
In this section we introduce our notation and conventions. We consider a 2D Bravais lattice with primitive vectors $\vec{a}_1$ and $\vec{a}_2$, see Fig.~1 of the main part of the manuscript. Accordingly, we label the unit cells with an integer vector $\vec{j}=(j_1,j_2)$. Then, we can write a generic tight-binding Hamiltonian as
\begin{align*}
\hat{H}&=\sum_{\vec{i},\vec{j}}\sum_{\alpha,\beta}t_{\vec{j}}^{\alpha,\beta}f^\dagger_{\vec{i},\alpha}f_{\vec{i}+\vec{j},\beta},
\end{align*}
where $f^\dagger_{\vec{i},\alpha}$ is the creation operator corresponding to an electron in unit-cell $\vec{i}$, and the index $\alpha$, which runs from $1$ to $N$ refers to the electronic internal degrees of freedom. It may therefore correspond to a spin, a sublattice or an orbital index. For example, in the honeycomb lattice model discussed in the main part of the manuscript it refers to the sublattice index. The choice of the unit cell is fixed by the dislocation under consideration, see for example Fig.~3 where the green rectangle denotes a preferred unit cell. To exploit the translation symmetry, we introduce the Fourier transformed creation and annihilation operators
\begin{align*}
f^\dagger_{\vec{q},\alpha}&:=\sum_{l_1=1}^{L_1}\sum_{l_2=1}^{L_2}e^{i\vec{q}\cdot\vec{l}}f^\dagger_{\vec{l},\alpha}/\sqrt{L_1 L_2}.
\end{align*} 
Note that $\vec{q}\cdot\vec{l}=(q_1\vec{b}_1+q_2\vec{b}_2)\cdot(l_1\vec{a}_1+l_2\vec{a}_2)/(2\pi)$, where $\vec{b}_1$ and $\vec{b}_2$ denote the reciprocal lattice vectors. Using these operators we can rewrite the Hamiltonian as
\begin{align*}
\hat{H}&=\sum_{\vec{q}\in BZ}\hat{H}(\vec{q})=\sum_{\alpha,\beta}\sum_{\vec{q}\in BZ}f^\dagger_{\vec{q},\alpha}\tilde{H}^{\alpha,\beta}(\vec{q})f_{\vec{q},\beta},
\end{align*}
where $\tilde{H}^{\alpha,\beta}(\vec{q})=\sum_{\vec{j}}t_{\vec{j}}^{\alpha,\beta}e^{i\vec{q}\cdot\vec{j}}$. We refer to $\hat{H}(\vec{q})$ as the second-quantized Hamiltonian, while $\tilde{H}(q)$ is its first quantized counterpart. 
We further denote the eigenstates of the first quantized Hamiltonian with $|\Psi_{n}(\vec{q})\rangle=[\Psi_{n,1}(\vec{q}),\ldots,\Psi_{n,N}(\vec{q})]^T$, where $n=1,\ldots,N$ is the band index. 
The real-space wave function with crystal momentum $q$ and band index $n$ within a given unit cell is proportional to $|\Psi_n(q)\rangle$. Finally, we need to define the Berry connection
\begin{align*}
\mathcal{A}^{m,n}_j(\vec{q})&=\langle\Psi_m(\vec{q})|i\partial_{q_j}|\Psi_n(\vec{q})\rangle,
\end{align*}
with $m,n=1,\ldots,N_F$, and the Abelian part of the Berry curvature, which is given by
\begin{align*}
\mathcal{F}_{i,j}=\partial_{q_i}\Tr{\mathcal{A}_i(\vec{q})}-\partial_{q_j}\Tr{\mathcal{A}_j(\vec{q})}
\end{align*}
The Berry phase $\gamma_\lambda$ correspondig to a generic contour $\lambda$ in the 2D Brillouin zone is defined as
\begin{align*}
\gamma_\lambda&=\int_\lambda\mathrm{d}\vec{q}\cdot\Tr\left[\vec{\mathcal{A}}(\vec{q})\right].
\end{align*}
If the contour is time-reversal symmetric, and $\hat{T}^2=-1$, then one can consider the partial Berry phase $\gamma^I_\lambda$, which is defined as
\begin{align*}
\gamma_\lambda^I&=\int_{\lambda^{1/2}}\mathrm{d}\vec{q}\cdot\Tr\left[\vec{\mathcal{A}}(\vec{q})\right],
\end{align*}
here $\lambda^{1/2}$ denotes half of the contour $\lambda$, and the union with the time-reversal copy, makes up the full contour $\lambda$. Moreover, it is required that one evaluates this contour integral using a time-reversal symmetric gauge: $|\Psi_n^I(\vec{q}\rangle=\tilde{T}|\Psi_n^{II}(-\vec{q})\rangle$, see also Refs.~\onlinecite{SMPRB76_045302,SMPRB96_235130}. In principle one may drop this constraint, by writing
\begin{align*}
\gamma_\lambda^I&=\int_{\lambda^{1/2}}\mathrm{d}\vec{q}\cdot\Tr\left[\vec{\mathcal{A}}(\vec{q})\right]+i \log\left(\frac{\Pf \mathcal{S}_{\tilde{T}}(\lambda^{1/2}_f)}{\Pf\mathcal{S}_{\tilde{T}}(\lambda^{1/2}_i)}\right),
\end{align*}
where $\lambda_{i(f)}^{1/2}$ denotes the time-reversal invariant starting (end) point of the contour $\lambda^{1/2}$.
\section{Detailed derivations of the main results}
Below we provide detailed derivations of the main results.
\subsection{Derivation of Eq.~(5)}
As discussed in the main text, we can express the dislocation charge in terms of the left and right edge charge of the two cylinders shown in Fig.~1 $Q=-Q_L-Q_R$. Hence, we simply need to calculate both separately. We note that both edges shown in Fig.~2 of the main part of the manuscript are translation symmetric in the direction $\vec{a}_1$. Hence, we may write
\begin{align*}
Q_L&=\sum_{j=0}^{L_1-1}Q_L\left(\frac{j2\pi}{L_1}\right)\quad\textrm{and}\quad Q_R=\sum_{j=0}^{L_1-2}Q_R\left(\frac{j2\pi}{L_1-1}\right),
\end{align*}
where $Q_{L(R)}(q_1)$ denotes the left (right) edge charge for a 1D insulator governed by the 1D Hamiltonian $\hat{H}_{q_1}(q_2):=\hat{H}(q_1,q_2)$. In Refs.~\onlinecite{SMPRB95_035421,SMPRB96_235130}, it has been shown that for a 1D crystalline insulator the right (left) edge charge is given by $+(-)\gamma/(2\pi)$, with 
\begin{align*}
\gamma&=\sum_{n\leq N_F}\int_0^{2\pi}\mathrm{d}q\langle\Psi_n(q)|i\partial_q|\Psi_n(q)\rangle.
\end{align*}
Therefore, we find
\begin{align*}
Q_L&=-\frac{1}{2\pi}\sum_{j=0}^{L_1-1}\gamma_2\left(\frac{j 2\pi}{L_1}\right)\quad\textrm{and}\quad Q_R=\frac{1}{2\pi}\sum_{j=0}^{L_1-2}\gamma_2\left(\frac{j 2\pi}{L_1-1}\right)
\end{align*}
with
\begin{align*}
\gamma_2\left(q_1\right)&=\int_0^{2\pi}\mathrm{d}q_2\langle\Psi_n(q_1,q_2)|i\partial_{q_2}|\Psi_n(q_1,q_2)\rangle=\int_0^{2\pi}\mathrm{d}q_2\Tr{\left[\mathcal{A}_2\left(\vec{q}\right)\right]}.
\end{align*}
As a result, we obtain the following expression for the dislocation charge:
\begin{align*}
Q&=\frac{1}{2\pi}\sum_{j=0}^{L_1-1}\gamma_2\left(\frac{j 2\pi}{L_1}\right)-\frac{1}{2\pi}\sum_{j=0}^{L_1-2}\gamma_2\left(\frac{j 2\pi}{L_1-1}\right)=\frac{1}{2\pi}\sum_{j=0}^{L_1-1}\left[\gamma_2\left(\frac{j 2\pi}{L_1}\right)-\gamma_2\left(\frac{j 2\pi}{L_1-1}\right)\right]+\frac{1}{2\pi}\gamma_2(2\pi)
\end{align*}
To make further progress, we use that $\gamma_2(q+\delta q)-\gamma_2(q)\approx\gamma_2'(q)\delta q$. We then find:
\begin{align*}
Q&\approx\frac{1}{2\pi}\sum_{j=0}^{L_1-1}\gamma'_2(\frac{j 2\pi}{L_1})\frac{j 2\pi}{L_1}\frac{-1}{L_1-1}+\frac{1}{2\pi}\gamma_2(2\pi)\approx-\frac{1}{(2\pi)^2}\int_0^{2\pi}\mathrm{d}q_1\gamma'_2(q_1)q_1+\frac{1}{2\pi}\gamma_2(2\pi)=\frac{1}{(2\pi)^2}\int_0^{2\pi}\mathrm{d}q_1\gamma_2(q_1)
\end{align*}
In the final line, we have used partial integration. In the limit, $L_1\rightarrow\infty$ the approximations become exact.
\subsection{Dislocation charge of time-reversal symmetric insulator}
In the presence of time-reversal symmetry, with $\hat{T}^2=-1$, we can refine the argument presented above. Indeed, time-reversal symmetry ensures that $Q_{L(R)}(q_2)=Q_{L(R)}(-q_2)$. For the moment, let us assume that $L_1$ is even. Then, we find that the total left edge charge is given by
\begin{align*}
Q_\textrm{L}&=2\sum_{j=1}^{L_1/2-1}Q_L\left(\frac{j2\pi}{L_1}\right)+Q_L(0)+Q_L(\pi)=-\frac{1}{\pi}\sum_{j=1}^{L_1/2-1}\gamma_2\left(\frac{j2\pi}{L_1}\right)+Q_L(0)+Q_L(\pi)
\end{align*}
For $q_1=0,\pi$, we find that the 1D Hamiltonian $\hat{H}_{q_1}$ is time-reversal symmetric. Hence, the corresponding edge charge is well-defined modulo $2$, see Ref.~\onlinecite{SMPRB96_235130} and can be expressed in terms of the partial Berry phase
\begin{align*}
Q_L(0)&=-Q_R(0)=-\frac{\gamma_2^I(0)}{\pi}=\frac{1}{\pi}\left[\int_0^\pi\mathrm{d}q_2\Tr\left[\mathcal{A}_2(0,q_2)\right]+i \log\left(\frac{\Pf \mathcal{S}_{\tilde{T}}(0,\pi)}{\Pf\mathcal{S}_{\tilde{T}}(0,0)}\right)\right]\textrm{ and }\\
Q_L(\pi)&=-\frac{\gamma_2^I(\pi)}{\pi}=\frac{1}{\pi}\left[\int_0^\pi\mathrm{d}q_2\Tr\left[\mathcal{A}_2(\pi,q_2)\right]+i \log\left(\frac{\Pf \mathcal{S}_{\tilde{T}}(\pi,\pi)}{\Pf\mathcal{S}_{\tilde{T}}(\pi,0)}\right)\right]
\end{align*}
Therefore, we find that the total left and right edge charges are given by
\begin{align*}
Q_L=-\frac{1}{\pi}\sum_{j=1}^{L_1/2-1}\gamma_2\left(\frac{j2\pi}{L_1}\right)-\frac{\gamma_2^I(0)}{\pi}-\frac{\gamma_2^I(\pi)}{\pi}\textrm{ and }Q_R=\frac{1}{\pi}\sum_{j=1}^{L_1/2-1}\gamma_2\left(\frac{j2\pi}{L_1-1}\right)+\frac{\gamma_2^I(0)}{\pi}.
\end{align*}
From this it follows that the dislocation charge is given by
\begin{align*}
Q&=\frac{1}{\pi}\sum_{j=1}^{L_1/2-1}\left[\gamma_2\left(\frac{j2\pi}{L_1}\right)-\gamma_2\left(\frac{j2\pi}{L_1-1}\right)\right]+\frac{\gamma^I_2\left(\pi\right)}{\pi},
\end{align*}
Following the same steps as above, we end up with 
\begin{align*}
Q&=\frac{1}{2\pi^2}\int_0^\pi\mathrm{d}q_1\gamma_2\left(q_1\right)+\frac{\gamma^I_2\left(\pi\right)}{\pi}-\frac{\gamma_2\left(\pi\right)}{2\pi}
\end{align*}
\subsection{Useful identities}
Here, we state important identities. Let $I$ denote the inversion matrix, i.e. $I\vec{q}=-\vec{q}$. Then we find that for inversion-symmetric system the following identity holds:
\begin{align*}
|\Psi_m(I\vec{q})\rangle&=\sum_{n\leq N_F}\mathcal{S}^\dagger_{\tilde{I}_A}(\vec{q})^{m,n}\tilde{I}_A(\vec{q})|\Psi_n(\vec{q})\rangle,\textrm{ and } \langle\Psi_m(I\vec{q})|=\sum_{n\leq N_F}\langle\Psi_n(\vec{q})|\tilde{I}_A^\dagger(\vec{q})\mathcal{S}_{\tilde{I}_A}(\vec{q})^{n,m}.
\end{align*}
Let $R$ denote the rotation matrix corresponding to an anti-clockwise rotation of $2\pi/3$:
\begin{align*}
R&=\begin{pmatrix}
0&-1\\
1&-1
\end{pmatrix}.
\end{align*}
Then, for $C_3$-symmetric systems we find
\begin{align*}
|\Psi_m(R\vec{q})\rangle&=\sum_{n\leq N_F}\mathcal{S}^\dagger_{\tilde{C}_{3,A}}(\vec{q})^{m,n}\tilde{C}_{3,A}(\vec{q})|\Psi_n(\vec{q})\rangle,\textrm{ and } \langle\Psi_m(R\vec{q})|=\sum_{n\leq N_F}\langle\Psi_n(\vec{q})|\tilde{C}_{3,A}^\dagger(\vec{q})\mathcal{S}_{\tilde{C}_{3,A}}(\vec{q})^{n,m}.
\end{align*}
As states in the main text, we find that the inversion (two-fold rotation) operator  is generically given by
\begin{align*}
\tilde{I}_A(\vec{q})&=\begin{pmatrix}
\tilde{I}_\mathcal{A}&0&0&0\\0&\tilde{I}_\mathcal{B}e^{-iq_1}&0&0\\0&0&\tilde{I}_\mathcal{C}e^{-i(q_1+q_2)}&0\\0&0&0&\tilde{I}_\mathcal{D}e^{-iq_2}
\end{pmatrix}
\end{align*}
and for $C_3$-symmetric systems we obtain
\begin{align*}
\tilde{C}_{3,A}(\vec{q})&=\begin{pmatrix}
\tilde{C}_{3,\mathcal{A}}&0&0\\0&\tilde{C}_{3,\mathcal{B}}e^{-iq_2}&0\\0&0&\tilde{C}_{3,\mathcal{C}}e^{i(q_1-q_2)}
\end{pmatrix}
\end{align*}
As a result, we find
\begin{align*}
\tilde{I}_A(\vec{q})^\dagger i\partial_1\tilde{I}_A(\vec{q})&=\begin{pmatrix}
0&0&0&0\\0&1&0&0\\0&0&1&0\\0&0&0&0
\end{pmatrix}\textrm{; }\tilde{I}_A(\vec{q})^\dagger i\partial_2\tilde{I}_A(\vec{q})=\begin{pmatrix}
0&0&0&0\\0&0&0&0\\0&0&1&0\\0&0&0&1
\end{pmatrix}.
\end{align*}
and
\begin{align*}
\tilde{C}_{3,A}(\vec{q})^\dagger i\partial_1\tilde{C}_{3,A}(\vec{q})&=\begin{pmatrix}
0&0&0\\0&0&0\\0&0&-1
\end{pmatrix}\textrm{; }\tilde{C}_{3,A}(\vec{q})^\dagger i\partial_2\tilde{C}_{3,A}(\vec{q})=\begin{pmatrix}
0&0&0\\0&1&0\\0&0&1
\end{pmatrix}.
\end{align*}
The identities allow us to relate the Berry connections at $\vec{q}$ and $I\vec{q}$ through
\begin{align}\label{eqSM:Berryrelinv}
\Tr\left[\mathcal{A}(I\vec{q})\right]&=I^T\left[\Tr\left[\mathcal{A}(\vec{q})\right]+\vec{\rho}(\vec{q})+i\nabla_{\vec{q}}\log{\left(\det\mathcal{S}^\dagger_{\tilde{I}_A}(\vec{q})\right)}\right]
\end{align}
with $\vec{\rho}(\vec{q})=(\rho_\mathcal{B}(\vec{q})+\rho_\mathcal{C}(\vec{q}),\rho_\mathcal{C}(\vec{q})+\rho_\mathcal{D}(\vec{q}))$. In addition we find that the Berry curvature satisfy
\begin{align}\label{eqSM:Berrycurvrelinv}
\mathcal{F}_{1,2}(I\vec{q})=\mathcal{F}_{1,2}(\vec{q})+\nabla\times\vec{\rho}(\vec{q}).
\end{align}
For $C_3$-symmetric systems we find
\begin{align}\label{eqSM:Berryrelc3}
\Tr\left[\mathcal{A}(R\vec{q})\right]&=(R^T)^2\left[\Tr\left[\mathcal{A}(\vec{q})\right]+\vec{\rho}(\vec{q})+i\nabla_{\vec{q}}\log{\left(\det\mathcal{S}^\dagger_{\tilde{C}_{3,A}}(\vec{q})\right)}\right],
\end{align}
with $\vec{\rho}(\vec{q})=(-\rho_\mathcal{C}(\vec{q}),\rho_\mathcal{B}(\vec{q})+\rho_\mathcal{C}(\vec{q}))$. This implies the following identity:
\begin{align}\label{eqSM:Berrycurvrelc3}
\mathcal{F}_{1,2}(R\vec{q})=\mathcal{F}_{1,2}(\vec{q})+\nabla\times\vec{\rho}(\vec{q}).
\end{align}
\subsection{Derivation of Eq.~(1)}
To derive the equation that links the geometric Berry phase to the electronic charge densities and the topological invariant $\xi_{i,X_A}$, with $X=C_2/I$, we first rewrite:
\begin{align}
\bar{\gamma}_1&=\frac{1}{2\pi}\int_0^{2\pi}\mathrm{d}q_2\gamma_1(q_2)=\gamma_1(0)+\frac{1}{2\pi}\int_0^{2\pi}\mathrm{d}q_1\int_0^{2\pi}\mathrm{d}q_2\mathcal{F}_{1,2}(\vec{q})q_2\nonumber\\
&=\gamma_1(0)+\frac{1}{2\pi}\int_{-\pi}^{\pi}\mathrm{d}q_1\int_{-\pi}^{\pi}\mathrm{d}q_2\mathcal{F}_{1,2}(\vec{q})q_2+\int_{-\pi}^{\pi}\mathrm{d}q_1\int_{-\pi}^{0}\mathrm{d}q_2\mathcal{F}_{1,2}(\vec{q})=\gamma_1(\pi)+\frac{1}{2\pi}\int_{-\pi}^{\pi}\mathrm{d}q_1\int_{-\pi}^{\pi}\mathrm{d}q_2\mathcal{F}_{1,2}(\vec{q})q_2\label{eqSM:gm1}
\end{align}
The expression on the RHS is completely gauge-invariant, i.e., it naturally implements the continuity constraint on $\gamma_1(q_2)$. Moreover, the domain is inversion-symmetric. Hence, we can take advantage of the fact that the Berry curvature at $-\vec{q}$ and $\vec{q}$ are related through Eq.~\eqref{eqSM:Berrycurvrelinv}. Using this result we find
\begin{align*}
\bar{\gamma}_1&=\gamma_1(\pi)-\pi\left[\rho_\mathcal{B}+\rho_\mathcal{C}\right]+\frac{1}{2}\int_{-\pi}^\pi\mathrm{d}q_1\left[\rho_\mathcal{C}(q_1,\pi)+\rho_\mathcal{D}(q_1,\pi)\right]
\end{align*}
We can combine the first and last term using the results from Ref.~\onlinecite{SMPRB96_235130}. Then, we finally obtain
\begin{align*}
\frac{1}{2\pi}\int_0^{2\pi}\mathrm{d}q_2\gamma_1(q_2)&=i\log(\xi_{1,X_A})-\pi\left[\rho_\mathcal{B}+\rho_\mathcal{C}\right].
\end{align*}
\subsection{Derivation of Eq.~(2)}
Next, let us relate the invariant $\chi_{i,X_{A}}$, with $X=C_2/I$, to the partial Berry phase $\gamma^I_i$. Again, we express the partial Berry phase in terms of the Berry curvature, i.e.
\begin{align*}
\pi q_1^I&=\frac{1}{2\pi}\int_0^{\pi}\mathrm{d}q_2\gamma_1(q_2)+\gamma_1^I(\pi)-\gamma_1(\pi)/2=\frac{1}{2\pi}\int_0^{2\pi}\mathrm{d}q_1\int_0^{\pi}\mathrm{d}q_2\mathcal{F}_{1,2}(\vec{q})q_2+\gamma_1^I(\pi)=\frac{1}{4\pi}\int_{-\pi}^{\pi}\mathrm{d}q_1\int_{-\pi}^{\pi}\mathrm{d}q_2\mathcal{F}_{1,2}(\vec{q})q_2+\gamma_1^I(\pi),
\end{align*}
where $q_1^I$ is defined in the main text, and in the final equality we have used that $\mathcal{F}_{1,2}(\vec{q})=-\mathcal{F}_{1,2}(-\vec{q})$ in the presence of time-reversal symmetry. Again, making use of Eq.~\eqref{eqSM:Berrycurvrelinv} we, therefore, obtain
\begin{align*}
\pi q_1^I&=-\frac{\pi}{2}\left[\rho_\mathcal{B}+\rho_\mathcal{C}\right]+\frac{1}{4}\int_{0}^{2\pi}\mathrm{d}q_1\left[\rho_\mathcal{C}(q_1,\pi)+\rho_\mathcal{D}(q_1,\pi)\right]+\gamma_1^I(\pi)=i\log(\chi_{1,X_A})-\frac{\pi}{2}\left[\rho_\mathcal{B}+\rho_\mathcal{C}\right].
\end{align*}
The final equality is derived in Ref.~\onlinecite{SMPRB96_235130} .
\subsection{Relation between $W_{\mu_1-\mu_3}\left(\mathcal{S}_{C_{3,A}}\right)$ and $\det{(\mathcal{S}_{\tilde{C}_{3,A}}(K^\pm)})$}
Here, we show that $W_{\mu_1-\mu_3}\left(\mathcal{S}_{C_{3,A}}\right)$ can be expressed in terms of the determinant of the sewing matrix at the two high-symmetry points $K^\pm$. For this purpose we express the winding number in terms of the contours $a$, $b$, and $c$, shown in Fig.~4.
\begin{align*}
W_{\mu_1-\mu_3}\left(\mathcal{S}_{C_{3,A}}\right)&=W_{2a+b-c}\left(\mathcal{S}_{C_{3,A}}\right).
\end{align*}
Next, we use that
\begin{align*}
\nabla_{\vec{q}}\det{\mathcal{S}_{\tilde{C}_{3,A}}(\vec{q})\mathcal{S}_{\tilde{C}_{3,A}}(R\vec{q})\mathcal{S}_{\tilde{C}_{3,A}}(R^{2}\vec{q})}=0.
\end{align*}
This allows us to write
\begin{align*}
W_{b-c}\left(\mathcal{S}_{C_{3,A}}\right)&=W_a\left(\mathcal{S}_{C_{3,A}}\right),
\end{align*}
where we have used that $b=-R a$ and $c=R^2 a$. Therefore, we can express the $\mathbb{Z}_3$ invariant in terms of the eigenvalues at the of the rotation operator at the high-symmetry points $K^+$ and $K^-$.
\begin{align*}
W_{\mu_1-\mu_3}=3W_{a}=i\frac{3}{2\pi}\log{\det{(\mathcal{S}_{\tilde{C}_{3,A}}(K^+)}/\det{(\mathcal{S}_{\tilde{C}_{3,A}}(K^-)}}.
\end{align*}
\subsection{Relation between $W^{1/2}_{\mu_1-\mu_3}\left(\mathcal{S}_{C_{3,A}}\right)$ and $\det{(\mathcal{S}_{\tilde{C}_{3,A}}(K^+)})$}
The argument presented above can also be used to relate $W^{1/2}_{\mu_1-\mu_3}\left(\mathcal{S}_{C_{3,A}}\right)$ to the determinant of the sewing matrix at $\vec{q}=K^+$. For this purpose, we write:
\begin{align*}
W^{1/2}_{\mu_1-\mu_3}\left(\mathcal{S}_{C_{3,A}}\right)&=W^{1/2}_{2a+b-c}\left(\mathcal{S}_{C_{3,A}}\right).
\end{align*}
Then it follwos that 
\begin{align*}
W^{1/2}_{\mu_1-\mu_3}\left(\mathcal{S}_{C_{3,A}}\right)&=3W^{1/2}_{a}\left(\mathcal{S}_{C_{3,A}}\right)=i\frac{3}{2\pi}\log{\det{(\mathcal{S}_{\tilde{C}_{3,A}}(K^+))}/\det{(\mathcal{S}_{\tilde{C}_{3,A}}(0,\pi))}}=i\frac{3}{2\pi}\log{\det{(\mathcal{S}_{\tilde{C}_{3,A}}(K^+))}}.
\end{align*}
In the final equality we have made use of the fact that the determinant is identically equal to $1$ at momenta that are time-reversal symmetric.
\subsection{Derivation of Eq.~(3)}
In this section, we express $\bar{\gamma_1}$ and $\bar{\gamma_2}$ in terms of the invariant $\xi_{C_{3,A}}$. First, we will relate the integral of the Berry connection along the contour $\mu_1$ to $\xi_{C_{3,A}}$. Making use of Eq.~\eqref{eqSM:Berryrelc3}, we find
\begin{align*}
\gamma_{\mu_{i+1}}&=\int_{\mu_{i}}\mathrm{d}\left(R\vec{q}\right)\cdot\Tr{\left[\mathcal{A}(R\vec{q})\right]}=\int_{\mu_{i}}\mathrm{d}\vec{q}\cdot\Tr{\left[\mathcal{A}(\vec{q})\right]}+i\int_{\mu_i}\mathrm{d}\vec{q}\cdot\nabla_{\vec{q}}\log{\left(\det\mathcal{S}^\dagger_{C_{3,A}}(\vec{q})\right)}+\int_{\mu_{i}}\mathrm{d}\vec{q}\cdot\vec{\rho}(\vec{q})\\
&=\gamma_{\mu_i}-2\pi W_{\mu_i}\left(\mathcal{S}_{C_{3,A}}\right)+\int_{\mu_{i}}\mathrm{d}\vec{q}\cdot\vec{\rho}(\vec{q})
\end{align*}
As a result, we then obtain
\begin{align}\label{eqSM:gammal1}
\gamma_{\mu_1}&=\frac{1}{3}\left[\gamma_{\mu_1}+\gamma_{\mu_2}+2\pi W_{\mu_1}\left(\mathcal{S}_{C_{3,A}}\right)-\int_{\mu_{1}}\mathrm{d}\vec{q}\cdot\vec{\rho}(\vec{q})+\gamma_{\mu_3}-2\pi W_{\mu_3}\left(\mathcal{S}_{C_{3,A}}\right)+\int_{\mu_{3}}\mathrm{d}\vec{q}\cdot\vec{\rho}(\vec{q})\right]\nonumber\\
&=\frac{2\pi}{3}W_{\mu_1-\mu_3}\left(\mathcal{S}_{C_{3,A}}\right)-\frac{1}{3}\int_{\mu_{1}-\mu_3}\mathrm{d}\vec{q}\cdot\vec{\rho}(\vec{q})=i\log{\xi_{C_{3,A}}}-\frac{1}{3}\int_{\mu_{1}-\mu_3}\mathrm{d}\vec{q}\cdot\vec{\rho}(\vec{q})
\end{align}
Next, let us turn to the expression for $\bar{\gamma_1}$, Eq.~\eqref{eqSM:gm1} derived in Sec. D:
\begin{align*}
\bar{\gamma}_1&=\gamma_1(\pi)+\frac{1}{2\pi}\int_{-\pi}^{\pi}\mathrm{d}q_1\int_{-\pi}^{\pi}\mathrm{d}q_2\mathcal{F}_{1,2}(\vec{q})q_2
\end{align*}
\begin{figure}[b]
\centering
\includegraphics[width=.475\textwidth]{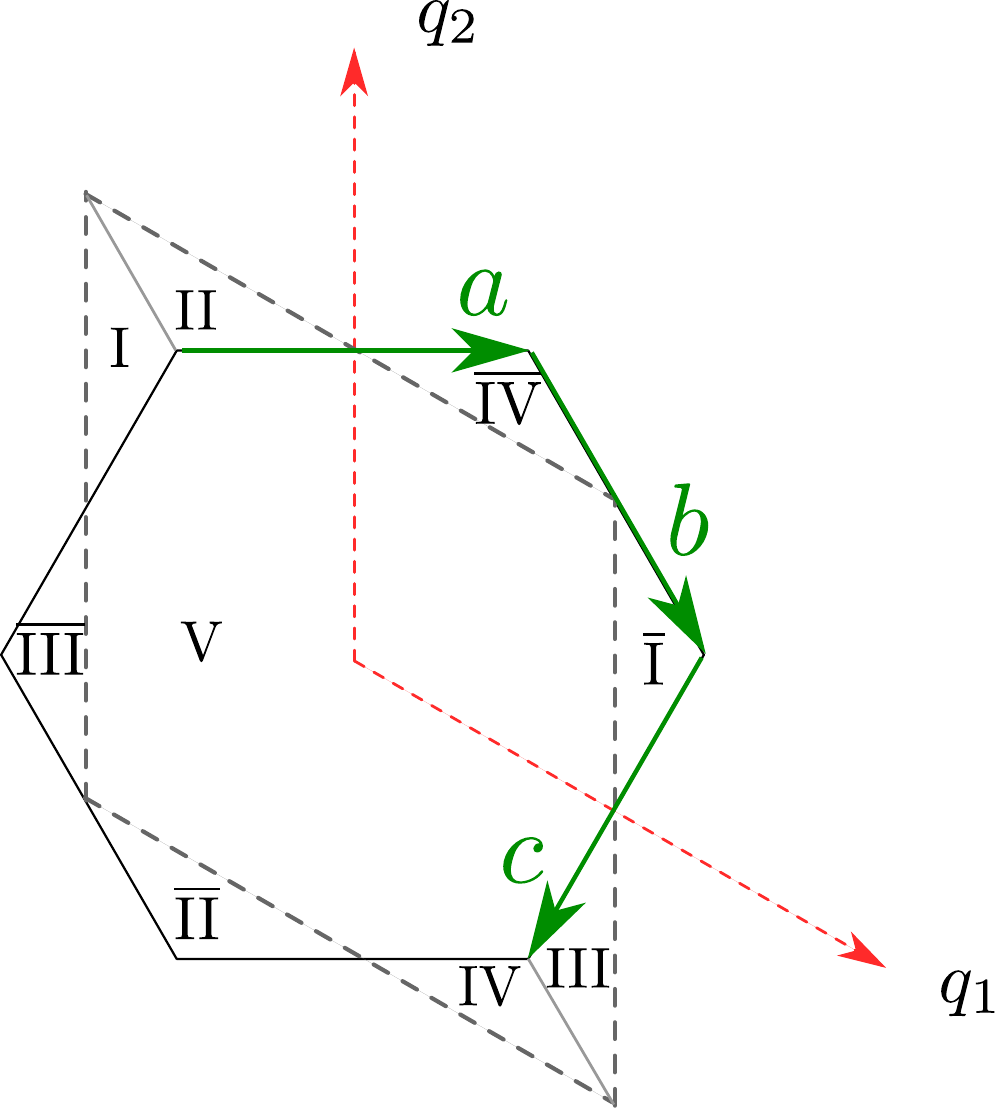}
\caption{The dashed lines enclose the domain $BZ=\left[-\pi,\pi\right]\times\left[-\pi,\pi\right]$. The green arrows denote the contours $a$, $b$, and $c$.}
\end{figure}
To exploit the rotational symmetry, we will change the domain of integration in the second term. For this purpose we divide the domain $[-\pi,\pi]\times[-\pi,\pi]$ into five parts, see Fig.~4, labeled as $I, II, III, IV$, and $V$, i.e.
\begin{align*}
BZ&:=[-\pi,\pi]\times[-\pi,\pi]=I\cup II\cup III\cup IV \cup V.
\end{align*}
We wish to express the integral over $BZ$ as an integral over the symmetric domain $\overline{BZ}$, see Fig.~4,  consisting of $\overline{I},\overline{II},\overline{III},\overline{IV}$, and $V$, i.e.
\begin{align*}
\overline{BZ}&:=\overline{I}\cup \overline{II}\cup \overline{III}\cup \overline{IV} \cup V.
\end{align*}
Using this notation, we can write
\begin{align}\label{eqSM:gamma1}
\bar{\gamma}_1&=\gamma_1(\pi)+\frac{1}{2\pi}\int_{\overline{BZ}}\mathrm{d}\vec{q}\mathcal{F}_{1,2}(\vec{q})q_2+\int_{\overline{II}-\overline{IV}}\mathrm{d}\vec{q}\mathcal{F}_{1,2}(\vec{q})=\gamma_{\mu_1}+\frac{1}{2\pi}\int_{\overline{BZ}}\mathrm{d}\vec{q}\mathcal{F}_{1,2}(\vec{q})q_2.
\end{align}
In the second equality we have employed Stokes' theorem. To make further progress, we need to specify the integration bounds explicitly. Here, we use the following parametrization:
\begin{align*}
\int_{\overline{BZ}}\mathrm{d}q_1\mathrm{d}q_2=\int_{-4\pi/3}^{4\pi/3}\mathrm{d}q_1\int_{b^{-}(q_1)}^{b^+(q_1)}\mathrm{d}q_2,
\end{align*}
with $b^+(q_1)$ given by
\begin{align*}
b^+(q_1)&=\begin{cases}2q_1+2\pi\quad&\textrm{if }-\frac{4\pi}{3}\leq q_1\leq -\frac{2\pi}{3}\\
\frac{1}{2}q_1+\pi\quad&\textrm{if }-\frac{2\pi}{3}\leq q_1\leq \frac{2\pi}{3}\\
-q_1+2\pi\quad&\textrm{if }\frac{2\pi}{3}\leq q_1\leq \frac{4\pi}{3}
\end{cases}
\end{align*}
and $b^{-}(q_1)$ given by
\begin{align*}
b^-(q_1)&=\begin{cases}-q_1-2\pi\quad&\textrm{if }-\frac{4\pi}{3}\leq q_1\leq -\frac{2\pi}{3}\\
\frac{1}{2}q_1-\pi\quad&\textrm{if }-\frac{2\pi}{3}\leq q_1\leq \frac{2\pi}{3}\\
2q_1-2\pi\quad&\textrm{if }\frac{2\pi}{3}\leq q_1\leq \frac{4\pi}{3}
\end{cases}
\end{align*}
Note that $b^+$ is related to $b^-$ in the following way
\begin{align}\label{eqSM:relb}
b^+(q_1)&=\begin{cases}b^-(q_1+2\pi) &\textrm{ if } -\frac{4\pi}{3}\leq q_1\leq-\frac{2\pi}{3}\\
b^-(q_1)+2\pi & \textrm{ if }  -\frac{2\pi}{3}\leq q_1\leq\frac{2\pi}{3}\\
b^-(q_1-2\pi) +2\pi& \textrm{ if }  \frac{2\pi}{3}\leq q_1\leq\frac{4\pi}{3}\end{cases}
\end{align}
With the help of Eq.~\eqref{eqSM:Berrycurvrelc3}, we can express the second term in Eq.~\eqref{eqSM:gamma1} as
\begin{align}
\label{eqSM:intF}
\int_{\overline{BZ}}\mathcal{F}_{1,2}(\vec{q})q_2&=\frac{1}{3}\int_{\overline{BZ}}\mathrm{d}q\left[\mathcal{F}_{1,2}(\vec{q})+\mathcal{F}_{1,2}(R\vec{q})+\mathcal{F}_{1,2}(R^2\vec{q})\right]q_2-\frac{1}{3}\int\mathrm{d}q_1\int\mathrm{d}q_2\left(\partial_{q_1}\rho_2(\vec{q})-\partial_{q_2}\rho_1(\vec{q})\right)q_2\nonumber\\
&-\frac{1}{3}\int\mathrm{d}q_1\int\mathrm{d}q_2\left(-\partial_{q_2}\rho_2(\vec{q})-\partial_{q_1}\rho_1(\vec{q})-\partial_{q_2}\rho_1(\vec{q})\right)q_2\nonumber\\
&=\underbrace{-\frac{1}{3}\int_{-4\pi/3}^{4\pi/3}\mathrm{d}q_1\int_{b^-(q_1)}^{b^+(q_1)}\mathrm{d}q_2[\partial_{q_1}\rho_2(\vec{q})]q_2}_\textrm{\eqref{eqSM:intF}a}+\underbrace{\frac{2}{3}\int_{-4\pi/3}^{4\pi/3}\mathrm{d}q_1\int_{b^-(q_1)}^{b^+(q_1)}\mathrm{d}q_2[\partial_{q_2}\rho_1(\vec{q})]q_2}_\textrm{\eqref{eqSM:intF}b}\nonumber\\
&+\underbrace{\frac{1}{3}\int_{-4\pi/3}^{4\pi/3}\mathrm{d}q_1\int_{b^-(q_1)}^{b^+(q_1)}\mathrm{d}q_2[\partial_{q_1}\rho_1(\vec{q})]q_2}_\textrm{\eqref{eqSM:intF}c}+\underbrace{\frac{1}{3}\int\mathrm{d}q_1\int\mathrm{d}q_2[\partial_{q_2}\rho_2(\vec{q})]q_2}_\textrm{\eqref{eqSM:intF}d}
\end{align}
Let us now work out the terms on the RHS of the equation above. Let us start with the  first term.  We move the derivative in front of the integral over $q_2$:
\begin{align*}
\eqref{eqSM:intF}\textrm{a}&=-\frac{1}{3}\int_{-4\pi/3}^{4\pi/3}\mathrm{d}q_1\partial_{q_1}\int_{b^-(q_1)}^{b^+(q_1)}\mathrm{d}q_2\rho_2(\vec{q})q_2+\frac{1}{3}\int_{-4\pi/3}^{4\pi/3}\mathrm{d}q_1b^+(q_1)\rho_2(q_1,b^+(q_1))\partial_{q_1}b^+(q_1)\nonumber\\
&-\frac{1}{3}\int_{-4\pi/3}^{4\pi/3}\mathrm{d}q_1b^-(q_1)\rho_2(q_1,b^-(q_1))\partial_{q_1}b^-(q_1)\nonumber
\end{align*}
The first term on the RHS vanishes because $b^+(\pm4\pi/3)=b^{-}(\pm4\pi/3)$.  With the help of Eq.~\eqref{eqSM:relb}, we can combine the second and third term:
\begin{align*}
\frac{1}{6}\int_{-2\pi/3}^{2\pi/3}\mathrm{d}q_1\rho_2(q_1,b^+(q_1))\cdot 2\pi+\frac{2}{3}\int_{-4\pi/3}^{-2\pi/3}\mathrm{d}q_1\rho_2(q_1,b^+(q_1))\cdot 0-\frac{1}{3}\int_{2\pi/3}^{4\pi/3}\mathrm{d}q_1\rho_2(q_1,b^+(q_1))\cdot2\pi=0.
\end{align*}
In the equation above the first and third term cancel against each other. Next, we consider the second term on the RHS of Eq.~\eqref{eqSM:intF}. Upon integration by parts, we find
\begin{align*}
\eqref{eqSM:intF}\textrm{b}&=-\frac{2}{3}\int\mathrm{d}q_1\int\mathrm{d}q_2\rho_1(\vec{q})+\frac{2}{3}\int_{-4\pi/3}^{4\pi/3}\mathrm{d}q_1[b^+(q_1)\rho_1(q_1,b^+(q_1))-b^-(q_1)\rho_1(q_1,b^-(q_1)]\nonumber\\
&=-\frac{2}{3}\int\mathrm{d}q_1\int\mathrm{d}q_2\rho_1(\vec{q})+\frac{4\pi}{3}\int_{-2\pi/3}^{4\pi/3}\mathrm{d}q_1\rho_1(q_1,b^+(q_1))=-\frac{8 \pi^2}{3}\rho_1+\frac{4\pi}{3}\int_{-2\pi/3}^{4\pi/3}\mathrm{d}q_1\rho_1(q_1,b^+(q_1)).
\end{align*}
The third term vanishes for the same reason as the first term:
\begin{align*}
\eqref{eqSM:intF}\textrm{c}&=\frac{1}{3}\int_{-4\pi/3}^{4\pi/3}\mathrm{d}q_1\partial_{q_1}\int_{b^-(q_1)}^{b^+(q_1)}\mathrm{d}q_2\rho_1(\vec{q})q_2\nonumber\\
&-\frac{1}{3}\int_{-4\pi/3}^{4\pi/3}\mathrm{d}q_1[b^+(q_1)\partial_{q_1}b^+(q_1)\rho_1(q_1,b^+(q_1))-b^-(q_1)\partial_{q_1}b^-(q_1)\rho_1(q_1,b^-(q_1))\nonumber\\
&=0
\end{align*}
Finally, let us rewrite the fourth term:
\begin{align*}
\eqref{eqSM:intF}\textrm{d}&=-\frac{1}{3}\int\mathrm{d}q_1\int\mathrm{d}q_2\rho_2(\vec{q})+\frac{1}{3}\int_{-4\pi/3}^{4\pi/3}\mathrm{d}q_1[b^+(q_1)\rho_2(q_1,b^+(q_1))-b^-(q_1)\rho_2(q_1,b^-(q_1)]\nonumber\\
&=-\frac{1}{3}\int\mathrm{d}q_1\int\mathrm{d}q_2\rho_2(\vec{q})+\frac{2\pi}{3}\int_{-2\pi/3}^{4\pi/3}\mathrm{d}q_1\rho_2(q_1,b^+(q_1))=-\frac{4\pi^2}{3}\rho_2+\frac{2\pi}{3}\int_{-2\pi/3}^{4\pi/3}\mathrm{d}q_1\rho_2(q_1,b^+(q_1))
\end{align*}
Hence, if we now combine Eqs.~\eqref{eqSM:gammal1}, \eqref{eqSM:gamma1}, and \eqref{eqSM:intF} we find
\begin{align*}
\bar{\gamma}_1&=i\log{\xi_{C_{3,A}}}-\frac{1}{3}\int_{\mu_{1}-\mu_3}\mathrm{d}\vec{q}\cdot\vec{\rho}(\vec{q})-\frac{4\pi}{3}\rho_1-\frac{2\pi}{3}\rho_2+\frac{1}{3}\int_{-2\pi/3}^{4\pi/3}\mathrm{d}q_1\rho_2(q_1,b^+(q_1))+\frac{2}{3}\int_{-2\pi/3}^{4\pi/3}\mathrm{d}q_1\rho_1(q_1,b^+(q_1))\\
&=i\log{\xi_{C_{3,A}}}-\frac{4\pi}{3}\rho_1-\frac{2\pi}{3}\rho_2=i\log{\xi_{C_{3,A}}}-\frac{2\pi}{3}\rho_\mathcal{B}+\frac{2\pi}{3}\rho_\mathcal{C}
\end{align*}
Here, we have used that the second term cancels the last two integrals. Following the same steps for $\gamma_2$, we obtain
\begin{align*}
\bar{\gamma}_2&=i\log{\xi_{C_{3,A}}}+\frac{2\pi}{3}\rho_1-\frac{2\pi}{3}\rho_2=i\log{\xi_{C_{3,A}}}-\frac{2\pi}{3}\rho_\mathcal{B}-\frac{4\pi}{3}\rho_\mathcal{C}
\end{align*}
\subsection{Derivation of Eq.~(4)}
First, let us relate the partial Berry phase along the contour $\mu_1$ to the $\mathbb{Z}_3$-invariant $\chi_{C_{3,A}}$. For the purpose of the proof, we assume that we have found a time-reversal symmetric gauge along the contours $\mu_i$. Then we can write
\begin{align*}
\gamma_{\mu_{i+1}}^I&=\int_{\mu_{i}^{1/2}}\mathrm{d}R\vec{q}\cdot\Tr{\left[\mathcal{A}(R\vec{q})\right]}=\int_{\mu_{i}^{1/2}}\mathrm{d}\vec{q}\cdot\Tr{\left[\mathcal{A}(\vec{q})\right]}+i\int_{\mu_i^{1/2}}\mathrm{d}\vec{q}\cdot\nabla_{\vec{q}}\log{\det{\mathcal{S}^\dagger_{\tilde{C}_{3,A}}(\vec{q})}}+\int_{\mu_i^{1/2}}\mathrm{d}\vec{q}\cdot\vec{\rho}(\vec{q})\\
&=\gamma^I_{\mu_i}-2\pi W_{\mu_i}^{1/2}(\mathcal{S}_{\tilde{R}_A})+\int_{\mu_i^{1/2}}\mathrm{d}\vec{q}\cdot\vec{\rho}(\vec{q}).
\end{align*}
This implies:
\begin{align}\label{eqSM:partialgammal1}
\gamma_{\mu_1}^I&=\frac{2\pi}{3}W^{1/2}_{\mu_1-\mu_3}(\mathcal{S}_{\tilde{R}_A})-\frac{1}{3}\int_{\mu_1^{1/2}-\mu_3^{1/2}}\mathrm{d}\vec{q}\cdot\vec{\rho}(\vec{q})=i\log{\chi_{C_{3,A}}}-\frac{1}{3}\int_{\mu_1^{1/2}-\mu_3^{1/2}}\mathrm{d}\vec{q}\cdot\vec{\rho}(\vec{q}\nonumber)\\
&=i\log{\chi_{C_{3,A}}}-\frac{1}{6}\int_{\mu_1-\mu_3}\mathrm{d}\vec{q}\cdot\vec{\rho}(\vec{q})
\end{align}
Now, let us consider $q_1^I$. First, we express $q_1^I$ as an integral over the rotation-symmetric domain $\overline{BZ}$
\begin{align*}
\pi q_1^I&=\gamma_1^I(\pi)+\frac{1}{4\pi}\int_{-\pi}^\pi\mathrm{d}q_1\int_{-\pi}^\pi\mathrm{d}q_2\mathcal{F}_{1,2}(\vec{q})q_2=\gamma_1^I(\pi)+\frac{1}{2}\int_{\overline{II}-\overline{IV}}\mathrm{d}\vec{q}\mathcal{F}_{1,2}(\vec{q})+\frac{1}{4\pi}\int_{\overline{BZ}}\mathrm{d}\vec{q}\mathcal{F}_{1,2}(\vec{q})q_2\\
&=\gamma_1^I(\pi)+\int_{\overline{II}}\mathrm{d}\vec{q}\mathcal{F}_{1,2}(\vec{q})+\frac{1}{4\pi}\int_{\overline{BZ}}\mathrm{d}\vec{q}\mathcal{F}_{1,2}(\vec{q})q_2=\gamma_{\mu_1}^I+\frac{1}{4\pi}\int_{\overline{BZ}}\mathrm{d}\vec{q}\mathcal{F}_{1,2}(\vec{q})q_2.
\end{align*}
In the third equality we have used that $\mathcal{F}_{1,2}(\vec{q})=-\mathcal{F}_{1,2}(-\vec{q})$, and in the fourth equality we have employed Stokes theorem. Using Eqs.~\eqref{eqSM:intF} and \eqref{eqSM:partialgammal1} we finally obtain
\begin{align*}
\pi q_1^I&=i\log{(\chi_{C_{3,A}})}-\frac{\pi}{3}\rho_\mathcal{B}+\frac{\pi}{3}\rho_\mathcal{C}
\end{align*}
In a similar fashion we find
\begin{align*}
\pi q_2^I&=i\log{(\chi_{C_{3,A}})}-\frac{\pi}{3}\rho_\mathcal{B}-\frac{2\pi}{3}\rho_\mathcal{C}.
\end{align*}
\subsection{Spin-orbit coupling Hamiltonians}
Here, we give the SOC Hamiltonians that we have used for the honeycomb lattice. The Rashba SOC Hamiltonian reads:
\begin{align*}
\tilde{H}_\textrm{RSO}(\vec{q})&=i\lambda_\textrm{RSO}\begin{pmatrix}
0&1\\
-1&0
\end{pmatrix}\otimes\left(\frac{\sqrt{3}}{2}\sigma_2+\frac{1}{2}\sigma_1\right)+i\lambda_\textrm{RSO}\begin{pmatrix}
0&e^{-i q_1}\\
-e^{i q_1}&0
\end{pmatrix}\otimes\left(-\frac{\sqrt{3}}{2}\sigma_2+\frac{1}{2}\sigma_1\right)-i\lambda_\textrm{RSO}\begin{pmatrix}
0&e^{-iq_2}\\
-e^{iq_2}&0
\end{pmatrix}\otimes\sigma_1,
\end{align*}
and
\begin{align*}
\tilde{H}_\textrm{ISO}(\vec{q})&=\lambda_\textrm{ISO}f(\vec{q})\begin{pmatrix}
1&0\\0&-1
\end{pmatrix}\otimes\sigma_3.
\end{align*}

\end{widetext}

\begin{thebibliography}{10}
\bibitem{intercellular}
\textrm{The Berry phase $\gamma_{1/2}(q_{2/1})$ should be calculated using the periodic gauge, i.e. $|\Psi_n(\vec{q})\rangle=|\Psi_n(\vec{q}+\vec{b}_{1/2})\rangle$, where $\vec{b}_{1/2}$ denotes a reciprocal lattice vector. This Berry phase corresponds to the inter-cellular part of the Zak phase, see Refs.~\cite{PRB96_235130,PRB95_035421}.}
\bibitem{z2}
\textrm{For completeness we note that one may drop the time-reversal symmetry constraint, by writing
\unexpanded{\begin{align*}
~W^{1/2}_{\lambda_1}(\mathcal{S}_{\tilde{I}(\tilde{C_2})_A})=\frac{i}{2\pi}\left[\int_0^\pi dq\partial_q\log{(\det{\mathcal{S}_{\tilde{I}(\tilde{C_2})_A}(q,\pi)})}\right.\nonumber\\
\left.+\int_0^\pi dq\partial_q\log{(\det{\mathcal{S}_{\tilde{T}}(q,\pi)})}-2\log\left(\frac{\Pf{\mathcal{S}_{\tilde{T}}(\pi,\pi)}}{\Pf{\mathcal{S}_{\tilde{T}}(0,\pi)}}\right)\right],
\end{align*}}
and likewise for  $W^{1/2}_{\lambda_2}(\mathcal{S}_{\tilde{I}(\tilde{C_2})_A})$. If a time-reversal symmetric gauge is employed the last two terms on the r.h.s. of the equation above vanish, and under an arbitrary gauge transformation the sum of the three terms can only change by an integer multiple of $2$.}
\end{thebibliography}
\end{document}